# Inhomogeneous Coherent States in Small-World Networks: Application to the Brain Networks

**Bahruz Gadjiev*, Tatiana Progulova**

Institute of System Analysis and Management, Dubna State University, Dubna, Russia

**Email address: bahruz_gadjiev@yahoo.com**

**Abstract:** We study the dynamics of the processes in the small-world networks with a power-law degree distribution where every node is considered to be in one of the two available statuses. We present an algorithm for generation of such network and determine analytically a temporal dependence of the network nodes degrees and using the maximum entropy principle we define a degree distribution of the network. We discuss the results of the Ising discrete model for small-world networks and in the framework of the continuous approach using the principle of least action, we derive an equation of motion for the order parameter in these networks in the form of a fractional differential equation. The obtained equation enables the description of the problem of a spontaneous symmetry breaking in the system and determination of the spatio-temporal dependencies of the order parameter in varies stable phases of the system. In the cases of one and two component order parameters with taken into account major and secondary order parameters we obtain analytical solutions of the equation of motion for the order parameters and determine solutions for various regimes of the system functioning. We apply the obtained results to the description of the processes in the brain and discuss the problems of emergence of mind.

## 1. Introduction

The macroscopic data of the functional imaging such as a functional magnetic resonance imaging and electroencephalography demonstrate a collective activity of many brain neurons. These investigations suggest that cognition and perception are the products of the collective activity of the neurons inside a large-scale brain network [1, 2].

A human brain consists of $\sim 10^{10}$ neurons, each of them is connected with the other neurons by $\sim 10^4$ links. The question is how these neurons acting in a coordinated manner form a coherent state, the time evolution of which we call thinking [2, 3].

Recently Eguiluz et al. [4] have presented a method of construction of functional brain networks proceeding from the results of the functional magnetic resonance imaging measurements in a human. In these experiments, a magnetic resonance activity of certain parts of the brain (so-called voxels) is measured at each discrete time step. By $x_i(t)$ we denote a voxel's activity at the instant of time $t$. It was proposed to consider that two voxels are functionally linked if the value of their temporal correlation exceeds a certain positive value $r_c$ independent of the value of their anatomical connection. The correlation coefficient between any pairs of voxels $x_i$ and $x_j$ is calculated as

$$r(i,j) = \frac{\langle x_i(t) x_j(t) \rangle - \langle x_i(t) \rangle \langle x_j(t) \rangle}{\sigma(x_i(t))\sigma(x_j(t))}, \qquad (1)$$

where $\sigma^2(x_l(t)) = \langle x_l^2(t) \rangle - \langle x_l(t) \rangle^2$, brackets $\langle ... \rangle$ represents temporal averages and $x_l(t)$ is the blood oxygenation level dependent on a signal of the voxel $i$ in case of the brain scanning data. The elements of the correlation matrix determine the value of correlations among various parts of the cerebral cortex. Using highly correlated nodes Eguiluz et al. have constructed a network and determined that the degree distribution of the obtained network has the form $p(k) \sim k^{-\gamma}$, where $\gamma \approx 2$ [4]. It has been also shown that these networks possess a small-world structure, a community structure and are fractals [5, 6].

The brain must surely functionate in a critical state as the states of the system near the critical regime are extremely sensitive to small actions. The size of the minimum area having sufficient features of a macroscopic system is called a correlation length. The correlation length of the system near the critical point is large enough. In this case the system is characterized by the fact that the number of the elements of the system in the area with the size about a correlation length tends to the infinity. So the behaviour of the degrees of freedom is interconnected and the behaviour of the system is defined by the fact of the coordinated behaviour of the degrees of freedom as well as by their nature.

The scale-invariant small-world network in the simplest case can be generated by the following algorithm being a generalization of the algorithm of the small-world network first proposed by Watts and Strogatz [7].

We proceed from a closed system of $N$ nodes with the periodic boundary conditions where each node is connected with the neighbors by $n$ links. A new edge is added to this system at each instant of time, one of the ends of this edge being connected with one of the nodes of the regular lattice with the probability $1/N$ while the other end of the edge being connected with that in accordance with the preferential attachment principle $k_i(t)/\sum_j k_j(t)$. Thus the change of the node degree $k_i(t)$ (without taking into account the contribution of the initial regular graph edges) is determined by two contributions and represented by the equation

$$\frac{\partial k_i(t)}{\partial t} = \frac{A}{N} + A \frac{k_i(t)}{\sum_j k_j(t)}. \quad (2)$$

Taking into account that at the instant $t$ a full network connectivity is $\sum_j k_j = 2t$ and that the change in the total degree of the network at one time step is $\Delta k = \sum_i (k_i(t) - k_i(t-1)) = 2$ we obtain $A = 1$. Then (2) takes the form

$$\frac{\partial k_i(t)}{\partial t} = \frac{1}{N} + \frac{k_i(t)}{\sum_j k_j(t)}. \quad (3)$$

The solution of this equation has the form

$$k_i(t) = 2t/N + ct^{1/2}, \quad (4)$$

where $c$ is the constant of the integration which can be determined from the condition $\sum_i k_i(t) = 2t$. The degree distribution of similar networks is conveniently derived using the maximum entropy principle in the presence of constraints $\langle k \rangle = k_0$ and $\langle k^2 \rangle = \mu_0$, which leads to

$$p_{st}(k) = Z^{-1} \left(1 - \frac{1}{\beta}(1-q)(k-k_0)^2\right)^{\frac{q}{1-q}}. \quad (5)$$

Here $q$ is a measure of complexity of the system. The value $q = 1$ corresponds to the degree distribution in the form of the Gaussian distribution. At $q \neq 1$ and for large enough $k$ the network is characterized by the power-law degree distribution [8]. The presented algorithm shows that order and disorder are inherent in small-world systems.

Discuss the problem of transition from the large world to the small one in networks. Rather, a smooth crossover is realized as the number of shortcuts grows. The small world within the limit of the infinite number of nodes $N$ is supposed to be obtained if $p = N_s/N$ is finite where $N_s$ is a number of shortcuts. If $p = 0$ we have a large world, that is a lattice [9]. It intuitively makes sense that in case of small enough, but finite $p$ the obtained small-world network could be considered as a lattice with a weak disorder. Then knowing the number of the components of the order parameter we could determine its transformational properties [10].

The analysis of the one-dimensional homogeneous Ising model with $N \to \infty$ vertices demonstrates the absence of a phase transition in the system at finite temperatures [11]. The phase transition emerges in the Katz model where apart from the short-range interactions there present weak interactions among all pairs of spins [9]. The one-dimensional Ising model has been studied in a series of works. It has been shown that a phase transition emerges at finite temperatures $T_c(p)$ for any finite $p$ [11].

In two-dimensional and three-dimensional lattices the Ising model shows a phase transition at finite temperatures [11]. Investigation of the role of the weak disorder in such systems demonstrates emergence of a phase transition with a change of the universality class [12]. Such results were obtained in a series of the investigations of the Ising model on small-world networks. Due to fluctuation of the correlations among the nodes the order parameter becomes a function weakly changing in space. In this case a spin interaction in the functional of free energy density can be represented as

$$\tfrac{1}{2}\sum_{i,j}\varphi_i\varphi_j \to \tfrac{1}{2}\int d^D x \left[\varphi^2(x) + (\nabla\varphi(x))^2\right].$$

Representation of a free energy in this way is typical for the Ginsburg-Landau theory, where the free energy according to transformation properties of the order parameter includes an integer basis of invariants.

Further, we introduce a free energy functional for small-world systems and using the principle of least action we derive an equation of motion for the order parameter representing spatial-temporal structures near the critical point.

## 2. Equation of Motion for the Order Parameter

To derive an equation of motion for the order parameter in the system with long-range space interactions and a temporal memory we determine the free energy functional $F[\eta]$ in the form: $F[\eta] = F_0[\eta] + F_I[\eta]$, where

$$F_0[\eta] = \int_R drdt \int_R dr'dt' \left\{\tfrac{1}{2}\tfrac{\partial\eta(r,t)}{\partial t}g_0(r,t,r',t')\tfrac{\partial\eta(r',t')}{\partial t'} + \tfrac{1}{2}\tfrac{\partial\eta(r,t)}{\partial r}g_1(r,t,r',t')\tfrac{\partial\eta(r',t')}{\partial r'}\right\}, \quad (6)$$

$$F_I[\eta] = \int_R drdt \int_R dr'dt' P(\eta(r,t),\eta(r',t'))V(\eta(r,t),\eta(r',t')). \quad (7)$$

Here $r$ is a spatial coordinate, $t$ is time and the functions $g_0(r,t,r',t')$ and $g_1(r,t,r',t')$ describe the influence of long-range space interactions and temporal memory on the dynamics of the order parameter. Integration is performed over the region $R$ in the two-dimensional space $R^2$ to which $(r,t)$ belong [8]. The dynamic equation follows from the stationary principle $\delta F[\eta, u] = 0$

$$\int_0^\infty dt' g(t,t')\tfrac{\partial\eta(r,t')}{\partial t'} + \int_0^\infty dr' k(r,r')\tfrac{\partial\eta(r',t)}{\partial r'} + \int_0^\infty d\eta(r',t') P(\eta(r,t),\eta(r',t'))\tfrac{\partial U(\eta(r,t),\eta(r',t'))}{\partial \eta(r',t')} = 0 \quad (8)$$

with separated spatial and temporal kernels, where $\delta F[\eta, u]$ is the Gateaux derivative. Considering the power-like kernels $g(t,t')$, $k(r,r')$ and $P(\eta(r,t),\eta(r',t'))$, we obtain a fractional differential equation for the order parameter as

$$g_0 {}_0^C D_t^\delta \eta(r,t) + k_0 {}_0^C D_r^v \eta(r,t) + {}_0^C D_{\eta(r,t)}^\mu U(\eta(r,t)) = 0, \quad (9)$$

where ${}_0^C D_t^\delta$ is the Caputo fractional derivative [12] and $1 < \delta \leq 2$, $1 < v \leq 2$, $0 < \mu \leq 1$. The term ${}_0^C D_{\eta(t,r)}^\mu U(\eta(r,t))$ consists of the non-integer powers of the order parameter. Note that choice of the fractional derivative depends on the type of the initial conditions and the processes, and other than Caputo derivative can appear. It is necessary to underline that the (9) is a generalization of the Ginzburg-Landau equation.

Further, we will discuss the solution of (9) in cases of one-component and two-component order parameters when the fractional derivatives in (9) are comfortable fractional derivatives [13]. Besides, we will also analyze the effects of secondary order parameters.

## 3. The Effects Induced by the Major One-Component Order Parameter

First, we take into account nonlinear equation (9) in the following form

$$\Phi\left(\frac{\partial^\alpha \eta}{\partial t^\alpha}, \frac{\partial^{2\alpha} \eta}{\partial t^{2\alpha}}, \frac{\partial^\beta \eta}{\partial t^\beta}, \frac{\partial^{2\beta} \eta}{\partial t^{2\beta}}, \dots\right) = 0. \tag{10}$$

Then, the following transformation applied: $\eta(r,t) = u(\xi)$. In order to use comfortable derivative we put

$$\xi = \frac{lx^\beta}{\beta} - \frac{\lambda t^\alpha}{\alpha}. \tag{11}$$

Using these transformations (10) we reduce the fractional differential equation to an integer nonlinear differential equation

$$\Psi(u(\xi), u'(\xi), u''(\xi), \dots) = 0. \tag{12}$$

According to (9) we consider the nonlinear fractional Klein–Gordon equation for the order parameter

$$\frac{\partial^{2\alpha} \eta(x,t)}{\partial t^{2\alpha}} - \frac{\partial^{2\beta} \eta(x,t)}{\partial x^{2\beta}} = -a\eta(x,t) - b\eta^3(x,t), \tag{13}$$

where $t > 0$ and $0 < \alpha \leq 1$, $0 < \beta \leq 1$. First we define the stationary solution of the equation

$$\frac{\partial^{2\beta} \eta(x,t)}{\partial x^{2\beta}} = a\eta(x,t) + b\eta^3(x,t). \tag{14}$$

We consider the case when $\beta = 1$. In this case the equation

$$\frac{\partial^2 \eta(x,t)}{\partial x^2} = a\eta(x,t) + b\eta^3(x,t) \tag{15}$$

is the Euler equation of the free energy functional

$$f(x) = a\eta^2 + \frac{b}{2}\eta^4 + \gamma\left(\frac{d\eta}{dx}\right)^2. \tag{16}$$

The homogeneous solutions $\eta^2 = \eta_{01} = 0$ and $\eta^2 = \eta_{02} = -a/b$ correspond to disordered and homogeneously ordered phases respectively.

After multiplying by $\eta'_x$ we obtain the first integral of (15)

$$\gamma(\eta')^2 = E + a\eta^2 + \frac{b}{2}\eta^4. \tag{17}$$

This equation admits separation of variables and the final solution of equation (17) has the form

$$\eta = \eta_1 sn\left(\gamma\eta_2 x \frac{b}{2}\right). \tag{18}$$

Now we determine the spatial distribution of the order parameter, which is described by fractional equation (14), and at $\alpha \to 1$ coincides with solution (18). Perform substitution of the variables $\eta(x,t) = u(\xi)$ and $\xi = lx^\beta/\beta$ and using (11) and (12) we obtain

$$l^2 u''_\xi = au(\xi) + bu^3(\xi). \tag{19}$$

Equation (19) with an accuracy of the notations coincides with (15) and consequently its solution is as follows

$$\eta = \eta_1 sn\left(\gamma\frac{b}{2}\eta_2\frac{lx^\beta}{\beta}\right). \tag{20}$$

The spatio-temporal distribution of the order parameter can be obtained through the transformation

$$\eta(x,t) = u(\xi) \text{ and } \xi = \frac{lx^\beta}{\beta} - \frac{\lambda t^\alpha}{\alpha}. \tag{21}$$

Using (11) and (12) equation (13) can be represented in the form

$$(l^2 + \lambda^2)u''_\xi = au(\xi) + bu^3(\xi). \tag{22}$$

Equations (19) and (22) with an accuracy of the notations coincide and consequently fractional differential equation (13) has a solution in the form

$$\eta(x,t) = \zeta_1 sn\left(\gamma\frac{b}{2}\zeta_2\left(\frac{lx^\beta}{\beta} - \frac{\lambda t^\alpha}{\alpha}\right)\right). \tag{23}$$

## 4. The Effects Induced by the Major Two-Component Order Parameter

According to (9) we consider the nonlinear fractional Sine-Gordon equation for the two-component order parameter

$$\frac{\partial^{2\alpha}\varphi(x,t)}{\partial t^{2\alpha}} - \frac{\partial^{2\beta}\varphi(x,t)}{\partial x^{2\beta}} = \sin\varphi. \tag{24}$$

First we determine a spatial distribution of the order parameter which is described by the equation

$$\frac{-d^2\varphi(x)}{dx^2} = \sin\varphi. \tag{25}$$

This equation is the Euler equation for the free energy functional in the form

$$f(x) = (\varphi')^2 + \cos\varphi. \tag{26}$$

After multiply by $\varphi'$ we obtain the first integral of this equation

$$(\varphi')^2 = 2(\cos\varphi + E). \tag{27}$$

This equation admits separation of variables and its solution has the form

$$\varphi(x) = 2am(cx, \kappa). \tag{28}$$

Now determine the spatial distribution of the order parameter which is described by the fractional equation

$$-\frac{\partial^{2\beta}\varphi(x,t)}{\partial x^{2\beta}} = \sin\varphi. \tag{29}$$

The substitution $\varphi(x,t) = u(\xi)$ and $\xi = lx^\beta/\beta$ leads fractional equation (24) to the ordinary differential equation

$$-l^2\frac{\partial^2\varphi(\xi)}{\partial \xi^2} = \sin\varphi(\xi). \tag{30}$$

Equations (25) and (30) with an accuracy of the notations coincide and consequently the solution of (29) has the form

$$\varphi(x) = 2am\left(c'\frac{lx^{\beta}}{\beta}, \kappa\right). \tag{31}$$

Equation (24) is solved through substitution (11) and (12) and has the form

$$\varphi(x,t) = 2am\left(c_0\left(\frac{lx^{\beta}}{\beta} - \frac{\lambda t^{\alpha}}{\alpha}\right), \kappa\right). \tag{32}$$

## 5. The Effects Induced by Secondary Order Parameters

Discuss the effects of secondary order parameters on the properties of the system. Consider the case of the one-component order parameter. If the transformational property of the secondary order parameter does not coincide with that of the major order parameter then the free energy functional describing the interaction between the major and secondary order parameters is represented by the expression

$$\varphi_{int} = c\eta^2\zeta^2 + \frac{e}{2}\zeta^4. \tag{33}$$

The equilibrium value of the secondary order parameter is determined from the equilibrium condition. Then we determine $\zeta^2 = 0$ and $\zeta^2 = -c\eta^2/e$. Substitution in the free energy functional leads to the renormalization $a' = a - c^2/(2e)$. So the spatial distribution of the secondary order parameter with an accuracy of the notations coincides with (18). Hence various phases with $\varphi(x,t) \neq 0$, $\zeta^2 = 0$ and $\varphi(x,t) \neq 0$, $\zeta^2 = -c\eta^2/e$ are possible.

In case of the two-component order parameter the free energy functional, describing the interaction between the major and secondary order parameters is represented by the expression

$$\varphi_{int} = k\varsigma \sin\frac{\varphi}{2} + \frac{g}{2}\varsigma^2, \tag{34}$$

and from the equilibrium condition we determine

$$\varsigma = -\frac{k}{g}\sin\frac{\varphi}{2}, \tag{35}$$

and consequently

$$\varphi_{int} = -\frac{k}{2g}\sin^2\left(\frac{\varphi}{2}\right) = \frac{k}{2g}(\cos\varphi - 1). \tag{36}$$

So the spatial distribution for the major order parameter with an accuracy of the notations has the form of (24) while the spatial distribution of the secondary order parameter is determined by relation (32).

## 6. Discussion

The main consequence of the Ginzburg-Landau theory is the possibility of enumeration of the phases manifested by the system. The brain consists of a great number of neurons (they may be considered as binary elements) interacting through synapses. In physics such system is known as ferromagnetic Ising model where Ising spins act as neurons and the binding energies (exchange integrals) correspond to the magnitude of synaptic bindings.

For free energy functional (16) we have homogeneous phases with $\eta^2 = \eta_{01} = 0$ and $\eta^2 = \eta_{02} = -a/b$ and a helicoidal phase with $\eta = \eta_1 sn(\gamma b\eta_2 x/2)$. There exist a disordered phase with $\eta^2 = \eta_{01} = 0$, an ordered magnetic phase with $\eta^2 = \eta_{02} = -a/b$ and an ordered helicoidal phase with $\eta = \eta_1 sn(\gamma b\eta_2 x/2)$ for the magnetic phase transition. The helicoidal phase is represented in the form of the alternating domain walls with opposite magnetic moments so that an average magnetization is equal to zero. With the change of the control parameter the distance between the soliton walls changes.

The origin of magnetism is a consequence of the fact that an electron with the charge $e$, mass $m_e$, and spin 1/2 has a magnetic moment $\pm e/(2m_e c)$. The Kadanov model or the method of the renormalization group and $\varepsilon$-expansion give us the details of the emergence of the coherent magnetically ordered phase when approaching the critical point [15]. In ferroelectrics a spontaneous polarization is an order parameter and the free energy functional leads to the emergence of a sequence of phase transitions high-symmetry – incommensurate – commensurate phase. Ordering of the dipole moments of the atomic complex gives rise to a ferroelectric phase. Interaction of the major order parameter with the secondary ones causes the emergence of additional phases and effects, for example, the phase with $\eta = \eta_1 \text{sn}(\gamma b \eta_2 x/2)$ and $\varsigma = 0$, and the phase with $\eta = \eta_1 \text{sn}(\gamma b \eta_2 x/2)$ and $\varsigma^2 = c\eta^2$.

By Haken's theory of self-organization the Ising model of ferromagnetics (or ferroelectrics) and the brain model belong to the same class of universality [16]. Hence it is interesting to discuss how physical properties of the brain, in particular mind can be explained from the point of view of physics.

The complexes possessing a dipole momentum emerge in the ferroelectrics. Their ordering causes a spontaneous polarization. Ordering of magnetic moments of electrons in magnets gives rise to a spontaneous magnetization. In book [3] developing the ideas of Alfred North Whitehead, Abner Shimony wrote about a possible presence of proto-mentality of elementary particles. In much the same way we can suppose that ratiomorphic behaviour inherent in living is explained by the presence of proto-mentality of particles. Then functioning of the brain can generate an ordering of proto-mentality of particles, ensuring the work of the body and as a secondary effect induces consciousness.

## 7. Conclusion

The network generated using the algorithm presented in this paper is a system with weak disorder. The regular substructure of this system possesses the symmetry of the discrete subgroup of the Galilean group [10]. To describe a concrete process in such system, it is necessary to construct the Rapp diagram, which allows determining the number of components of the order parameter [2, 10]. Thus, it is possible to define the integer basis of invariants of an irreducible representation on which depends the free energy functional [10]. The fractional differential equation of motion for the order parameter takes into account the effects of spatial and temporal memory. For multi-component order parameter, we have a system of coupled equations. In the present article, we obtained analytical solutions of fractional differential equations for the order parameter.

Both in one-component and two-component cases the brain functionates in the phases with a high response to an external perturbation and with a spatio-temporal distribution described by elliptic functions (32) and (36) respectively.

Thus a series of phases with an inhomogeneous spatio-temporal distribution of the order parameter depending on fractional degrees of time and a spatial coordinate emerge in small-world networks with a power-law degree distribution.

Prior to the creation of the microscopic theory of superconductivity of Bardeen – Cooper – Schrieffer, the phenomenological theory of superconductivity of Landau and Ginzburg was developed [16]. The theory of Landau and Ginzburg allowed to describe the features of measurable quantities near the change in the regimes of system functioning and create a theory of second-type superconductors. However, it was only after the creation of the microscopic theory of Bardeen – Cooper – Schrieffer superconductivity that the physical meaning of the order parameter was determined [16]. In our opinion, a similar situation takes place in the theory of brain functioning. The proposed approach can describe the properties of the system near the change of functioning regimes the brain and further, taking into account the ideas of Alfred North Whitehead, Abner Shimony, to determine the physical meaning of the order parameter.